\documentstyle[12pt]{article}
\newcommand{\be}{\begin{equation}}
\newcommand{\ee}{\end{equation}}
\newcommand{\ba}{\begin{eqnarray}}
\newcommand{\ea}{\end{eqnarray}}

\begin{document}
\begin{center}
{\bf\Huge  {Hamilton-Jacobi quantization
 of the finite dimensional  systems with constraints}}
\end{center} \begin{center} Dumitru Baleanu\footnote{ Permanent address
: Institute of Space Sciences, P.O.BOX, MG-23, R
76900, Magurele-Bucharest, Romania,E-Mail
address:~~baleanu@thsun1.jinr.ru, baleanu@venus.nipne.ro}\\ Bogoliubov
Laboratory of Theoretical Physics \\ Joint Institute for Nuclear
Research\\ Dubna, Moscow Region, Russia

\end{center}
\begin{center}
    and
\end{center}
\begin{center}
 Yurdahan $G\ddot{u}ler$ \footnote{E-Mail address:~~yurdahan@ari.cankaya.edu.tr}
\end{center}
\begin{center}
Department of Mathematics and Computer Sciences, Faculty of Arts and Sciences,
Cankaya University, Ankara , Turkey
\end{center}

\vskip 5mm
\bigskip
\nopagebreak
\begin{abstract}
 The Hamiltonian treatment of constrained systems in $G\ddot{u}ler's$
formalism  leads us to the total differential equations in many variables.
These equations are integrable if the corresponding system of
partial differential equations is a Jacobi system.
The main aim of this paper is to investigate the quantization of the
finite dimensional systems  with constraints using the canonical
formalism introduced by $G\ddot{u}ler$.
This approach is applied for two systems with
constraints and the results are in  agreement with those
obtained by Dirac's canonical  quatization method and path integral
quantization method. \end{abstract}

\section{Introduction}
 The canonical
formulation\cite{gu87},\cite{gu92},\cite{gu921},\cite{gu923},\cite{gu924},\cite{gu95}
 gives the set of Hamilton-Jacobi partial-differential equation as
\be H_{\alpha}^{'}(t_{\beta},q_{a},{\partial S\over\partial
q_{\alpha}},{\partial S\over\partial t_{\alpha}})=0,
\alpha,\beta=0,n-r+1,\cdots,n,a=1,\cdots,n-r,
\ee
where
\be
H_{\alpha}^{'}=H_{\alpha}(t_{\beta},q_{a},p_{a}) +p_{\alpha}
\ee
and $H_{0}$ are defined as
\be
H_{0}=-L(t,q_{i},{\dot q_{\nu}},{\dot q_{a}=w_{a}}) +p_{a}w_{a} +
{\dot q_{\mu}}p_{\mu}\mid_{p_{\nu}=-H_{\nu}},\nu=0,n-r+1,\cdots,n
\ee
 The equations of motion are obtained as total differential equations
in many variables as follows
\be\label{(pq)}
dq_{a}={\partial H_{\alpha}^{'}\over\partial p_{a}}dt_{\alpha},
dp_{a}=-{\partial H_{\alpha}^{'}\over\partial q_{a}}dt_{\alpha},
dp_{\mu}=-{\partial H_{\alpha}^{'}\over\partial t_{\mu}}dt_{\alpha},
\mu=1,\cdots, r
\ee

\be\label{(z)}
dz=(-H_{\alpha} +p_{a}{\partial H_{\alpha}^{'}\over\partial
p_{a}})dt_{\alpha} \ee where $z=S(t_{\alpha},q_{a})$.The set of
equations(\ref{(pq)},\ref{(z)}) is integrable if \be\label{(h)}
dH_{0}^{'}=0,dH_{\mu}^{'}=0,\mu=1,\cdots r
\ee
If conditions(\ref{(h)}) are not satisfied identically , one should consider
them as new constraints and again should test the consistency conditions.
Thus repeating this procedure one may obtain a new set of conditions.

The main aim of this paper is to investigate the quantization of
finite dimensional systems using $G\ddot{u}ler's$ formalism.
We test our formalism on two systems with first and second
class constraints respectively.

The plan of this paper is the following:

 A brief information of the $G\ddot{u}ler's$ formalism is given in
Sect.1.  In Sec.2 quantization of systems with constraints is
investigated.The examples are worked out
in Sec.3 In Sec.4 conclusions are presented.

\section{Quantization of the systems with constraints in
$G\ddot{u}ler's$ formalism}

Let us suppose that  for a finite dimensional system with constraints we
found all independent hamiltonians $H_{\mu}^{'}$  using the
calculus of variations\cite{gu87},\cite{gu92},\cite{gu923}. Because
all the hamiltonians $H_{\mu}^{'}$ are constraints we will use Dirac's
procedure of quatization\cite{Dirac}.
 We have
\be
H_{\mu}^{'}\Psi=0,\mu=1,\cdots,r \ee
where $\Psi$ is the wave
function.  The consistency conditions are \be\label{(cond)}
[H_{\mu}^{'},H_{\nu}^{'}]\Psi=0,\mu,\nu=1,\cdots r
\ee

If  for  a finite dimensional system  the hamiltonians $H_{\mu}^{'}$
satisfy
\be
[H_{\mu}^{'},H_{\nu}^{'}]=C_{\mu\nu}^{\alpha}H_{\alpha}^{'}
\ee
the system has first class constraints in the Dirac's classification.

On the other hand if
\be
[H_{\mu}^{'},H_{\nu}^{'}]=C_{\mu\nu}
\ee
where $C_{\mu\nu}$ do not depend of $q_{i}$ and $p_{i}$
 then from(\ref{(cond)})
we will be lead naturally to Dirac's brackets and the canonical
quatization will be performed taking Dirac's brackets into
commutators.

   $G\ddot{u}ler's$ formalism give us and the action when
all hamiltonians $H_{\mu}^{'}$  are in involution.   Because in this
formalism we work from the beginning in the extended space we suppose
that variables $t_{\alpha}$ depend of $\tau$.Here $\tau$ is canonical
conjugate to $p_{0}$.
We propose the following expression for the action

\be\label{(zet)}
z=\int(-H_{a} +p_{a}{\partial H_{a}^{'}\over\partial
p_{a}})\dot{t_{a}}{d\tau}
\ee
where $\dot{t_{\alpha}}={d t_{\alpha}\over\ d\tau}$.

If we are able , for a given
finite system with constraints, to find the independent hamiltonians
$H_{\mu}^{'}$ in involution then we can perform the quantization of
this system using path integral quantization method with the action
gives as in (\ref{(zet)}).

\section{Examples}
\subsection{A system with first class constraints}

Consider the following lagrangian(for more
details see\cite{gu92})
 \be\label{sist}
2L= a_{ij}{\dot q_{i}}{\dot q_{j}} +2b{\dot
q_{2}}-2c,i,j=1,2,3 \ee The generalized momenta read as \be
p_{1}=a_{1}{\dot q_{1}}, p_{2}=a_{2}({\dot q_{3}}-{\dot q_{2}}) +b,
p_{3}=a_{2}({\dot q_{3}}-{\dot q_{2}}) \ee Then we have two
hamiltonians in the $G\ddot{u}ler's$ formalism \be H^{'}_{0}= p_{0}
+{1\over 2}({p_{1}^{2}\over a_{1}}-{p_{3}^{2}\over a_{2}}) +c,
H^{'}_{2}=p_{2}+ p_{3}-b \ee
Here a,b and c are constants.

At this stage we have two
ways for quantization of the system presented above.

Because we have two constraints $H^{'}_{0}$ and $H^{'}_{2}$ Dirac's
canonical formalism for the systems with constraints will be
applied\cite{Dirac}.

Then we have
\be\label{(psi)}
H^{'}_{2}\Psi=0,H^{'}_{0}\Psi=0
\ee
where $\Psi$ is the wave function.

The consistency condition gives the following commutation relation
\be
[H^{'}_{2},H^{'}_{0}]\Psi=0
\ee
and it is automatically satisfied because
the hamiltonians $H^{'}_{2}$ and $H^{'}_{0}$ commute.

We found that a solution of eq.(\ref{(psi)})
has the following form
\be
\Psi=(-b^{2} +2c)e^{bq_{3}}[\sin{q_{1}} +\cos{q_{1}}]
\ee

For the path integral quantization we need the action z.
Using $G\ddot{u}ler's$ formalism we found
\be
dz=(-c + {p_{1}^{2}\over 2a_{1}} - {p_{3}^{2}\over 2a_{2}} +
b{\dot q_{2}})d\tau
\ee
or
\be\label{gul}
z=\int(-c + {p_{1}^{2}\over 2a_{1}} - {p_{3}^{2}\over 2a_{2}} +
b{\dot q_{2}})d\tau
\ee
We know that for a system with n degrees of freedom with r first
class-constraints $\phi^{a}$ the path integral representation is given
as[for more detailes see Ref.\cite{senj}]
\ba\label{(int)} &<q^{'}\mid exp[-i(t^{'}-t){\hat
H_{0}}]\mid
q>=&\int\prod_{t}d\mu(q_{j},p_{j})exp[i\{\int_{-\infty}^{+\infty}dt(p_{j}{\dot q_{j}}-H_{0})\}],\cr
&j=1,\cdots,n&
\ea
where the measure of integration is given as
\be
d\mu(q,p)=det\mid\{\psi^{a},\psi^{b}\}\mid\prod_{a=1}^{r}\delta(\chi^{a})\delta(\phi^{a})
\prod_{j=1}^{n}dq^{j}dp_{j}
\ee
and $\chi^{a}$ are r gauge constraints.

 If we perform now the usual
path integral quantization using (\ref{(int)}) for the system
(\ref{sist}) , after imposing the gauge
condition and integrate over $q_{2}$, we get the action (\ref{gul})
when $\tau$ is replaced by t.

\subsection{A system with  second class constraints}
Let us consider
the Lagrangian \cite{gu923}
\be\label{sist1} L={1\over 2}a_{1}{\dot q_{1}^{2}}-{1\over
2}a_{2}({\dot q_{2}^{2}} - 2{\dot q_{2}}{\dot q_{3}} +{\dot q_{3}^{2}}
+b{\dot q_{2}}-c \ee Here $a_{1}, a_{2}, b,c$ are functions of $q_{1},
q_{2},q_{3},t$. Let us specify the functions $a_{1},a_{2},b,c$ as
$a_{1}=1, a_{2}={1\over 2}, b=q_{1} + q_{3}, c=q_{1} +q_{2}
+q_{3}^{2}$.

From (\ref{sist1}) we found two hamiltonians $H_{0}^{'}$ and
$H_{2}^{'}$ as
\be\label{eq1} H_{0}^{'}=p_{0} +{1\over
2}(p_{1}^{2}-2p_{3}^{3}) +q_{1} +q_{2} + q_{3}^{2}=0 \ee \be\label{eq2}
H_{2}^{'}=p_{2} +p_{3}-q_{1}-q_{3}=0
\ee
If we impose the variations of(\ref{eq1}) and(\ref{eq2})
to be zero we get a new
independent hamiltonian
\be
H_{1}^{'}=-p_{1} +2p_{3} -2q_{3} -1=0
\ee
Then we have three independent hamiltonians $H_{0}^{'}$,$H_{1}^{'}$ and
$H_{2}^{'}$.

 Dirac's method of quantization give us the following
relations\cite{Dirac} \be H_{0}^{'}\Psi=0,  H_{1}^{'}\Psi=0,
H_{2}^{'}\Psi=0 \ee and the consistency conditions \be
[H_{0}^{'},H_{1}^{'}]\Psi=0
\ee
\be
[H_{0}^{'},H_{2}^{'}]\Psi=0
\ee
\be\label{pb12}
[H_{1}^{'},H_{2}^{'}]\Psi=0
\ee
Here $\Psi$ is the wave function.

 Because
\be
[H_{1}^{'},H_{2}^{'}]=1
\ee
and  the consistency condition(\ref{pb12}) is not satisfied.
In the Dirac's classification of constraints $H_{1}^{'}$ and
$H_{2}^{'}$ are second class constraints.
At this stage we will introduce  the Dirac's brackets for our system.

Some calculations gives the following form
\be
\{F,G\}_{D.B.}=\{F,G\} +\{F,H_{2}^{'}\}\{H_{1}^{'},G\}
-\{F,H_{1}^{'}\}\{H_{2}^{'},G\} \ee
and we can perform the canonical quantization taking Dirac's brackets
into commutators.

Now we would like to find the action using $G\ddot{u}ler's$ formalism.
Because the hamiltonians
$H_{0}^{'}$,$H_{1}^{'}$,$H_{2}^{'}$ are not in involution
 we extend the phase-space with another pairs of conjugate
variables $(\lambda, p_{\lambda})$ .The new  hamiltonians
$\tau_{1}^{'}$ ,$\tau_{2}^{'}$ and $\tau_{3}^{'}$ are in involution
and have the following expressions
\ba\label{c1}
&\tau_{1}^{'}&=p_{0} +{1\over 2}(p_{1}^{2}-2p_{3}^{3}) +q_{1} +q_{2} +
q_{3}^{2} +\lambda(-1 -4p_{3} +4q_{3})\cr
&+&p_{\lambda}(p_{1}-2p_{3}-1 +2q_{3})
 - {1\over 2}p_{\lambda}^{2}=p_{0} +\tau_{1}
\ea
\be\label{c2}
\tau_{2}^{'}=p_{2} +p_{3}-q_{1}-q_{3} +\lambda =p_{2}+ \tau_{2}
\ee
\be\label{c3}
\tau_{3}^{'}=-p_{1}-2q_{3} +2p_{3} -1-p_{\lambda}=-p_{\lambda}
+\tau_{3}
\ee

From(\ref{(z)}) we get the following expression for the action z
\be
dz=(-\tau_{1} +p_{1}{\dot q_{1}} +p_{2}{\dot q_{2}} +p_{3}{\dot q_{3}}
+p_{\lambda}{\dot\lambda} +{\dot q_{2}}p_{3} +{\dot\lambda}(-p_{1}+
2p_{3}))d\tau
\ee
or
\be\label{(action)}
z=\int(-\tau_{1} +p_{1}{\dot q_{1}} +p_{2}{\dot q_{2}} +p_{3}{\dot
q_{3}} +p_{\lambda}{\dot\lambda} +{\dot q_{2}}p_{3}
+{\dot\lambda}(-p_{1}+ 2p_{3}))d\tau
\ee
In this case the extended system has first class constraints in
Dirac's classification.
The action(\ref{(action)}) gives the same result as the
expression for the effective action from (\ref{(int)}) for the system
with first class constraints (\ref{c1}),(\ref{c2}),(\ref{c3}) when
we consider $t=\tau$.The only difference is that the gauge conditions
have different expressions in $G\ddot{u}ler's$ formalism.

\section{Concluding remarks}
In this paper  the quantization of the
finite dimensional systems  with constraints using  canonical
formalism introduced by $G\ddot{u}ler$ was investigated .

In  this formalism we have no first or
second class constraints as in the Dirac's classification at the
classical level  but the constraints arises naturally  in the set of
consistency conditions(\ref{(cond)}) at the quantum level.
The secondary constraints are obtained in $G\ddot{u}ler's$ formalism
using calculus of variations.
When a system has second class constraints
the Dirac's brackets  in $G\ddot{u}ler's$ formalism are defined  on the
extended space $(q_{i},\tau)$.
  In this case we enlarge the system and convert the second class
constraints into first class constraints in order to obtain
 the hamiltonians in involution.
  When the hamiltonians $H^{'}_{\mu}$ are in involution we can construct
the action in $G\ddot{u}ler's$ formalism.
 In this formalism all dependent variables $t_{\alpha}$ are
 gauge variables and we suppose that these variables have a dependence
 on $\tau$(a canonical conjugate variable with $p_{0}$).
In the case  when hamiltonians $H_{\mu}^{'}$ commute
  the action
(\ref{(zet)}) has the same form as obtained by path integral
 quantization method after performing all the calculations.
 In this
 case the results are in perfect agreement with those obtained by
 usual path integral quantization methods of a system with constraints.

  \section{Acknowledgements}

One of the authors (D.B.) would like to thank TUBITAK for financial
support  and METU for the hospitality during his
working stage at Department of Physics.


\begin{thebibliography}{99}
\bibitem{gu87}Y.$G\ddot{u}ler$,\sl{Il Nuovo Cimento B}, {\bf 100}
(1987) 251.
(1987) 267. 
\bibitem{gu92}Y.$G\ddot{u}ler$,\sl {Il Nuovo Cimento B}, {\bf
107} (1992) 1389.
 \bibitem{gu921}Y.$G\ddot{u}ler$,\sl{Il Nuovo Cimento
B}, {\bf 107} (1992) 1143. 
Cimento B, 113 (1998) 893. \bibitem{gu923} E.Rabei and
Y.$G\ddot{u}ler$,{\sl Phys.Rev.A},{\bf 46} (1992) 3513. 
\bibitem{gu924}
E.Rabei and Y.$G\ddot{u}ler$,\sl{Tr.J.of Physics} (1992) 297.
\bibitem{gu95}E.Rabei and Y.$G\ddot{u}ler$,\sl{Il Nuovo Cimento B},
{\bf 110} (1995) 893. 
\bibitem{Dirac}Dirac P.A.M, \sl{Lectures on Quantum
Mechanics } (Yeshiva University, New York, N.Y.)1964.
\bibitem{senj}Senjanovic P.,\sl{Ann.Phys.(N.Y.)},{\bf 100} (1976) 227.
\end{thebibliography}
\end{document}